\documentclass[aps,prl,reprint,superscriptaddress,nofootinbib]{revtex4-2}
\usepackage[T1]{fontenc}
\usepackage{amsmath,amssymb,bm,mathrsfs}
\usepackage{tikz,fancyhdr,shuffle}
\definecolor{pale}{HTML}{DCEEF0}
\usetikzlibrary{arrows.meta}
\usepackage{hyperref}
\DeclareFontFamily{U}{shuffle}{}
\DeclareFontShape{U}{shuffle}{m}{n}{%
  <3-8>shuffle7%
  <8->shuffle10%
}{}
\hypersetup{hidelinks,pdftitle={Universal Exact Splitting at Generic Kinematics}}
\newcommand{\Tr}{\operatorname{Tr}}
\newcommand{\A}{\mathcal A}
\newcommand{\J}{\mathcal J}
\newcommand{\F}{\mathcal F}
\newcommand{\mO}{\mathcal{O}}

\newcommand{\I}{\mathcal I}

\usepackage{xcolor}

\begin{document}

\title{Exact Splitting of Amplitudes and One-Loop Integrands: A Classical Origin of Zeros}

\author{Yang Li}
\email{leey0342@gmail.com}

\affiliation{Department of Physics, Technion -- Israel Institute of Technology, Haifa 3200003, Israel}
\author{Laurentiu Rodina}
\email{laurentiu.rodina@gmail.com}
\affiliation{Beijing Institute of Mathematical Sciences and
Applications (BIMSA), Beijing 101408, China}

\author{Zhi-Zhen Wang}
\email{wzzphy4869l@gmail.com}

\affiliation{Beijing Institute of Mathematical Sciences and
Applications (BIMSA), Beijing 101408, China}

\begin{abstract}
We introduce a recursion based on hidden zeros that reconstructs complete tree level scattering amplitudes and one-loop integrands at generic kinematics. The recursion takes a universal exact-splitting form: both tree amplitudes and one-loop surface integrands are expressed as sums of products of complete tree functions, with the chosen zero conditions manifest in every term. We establish this structure for cubic scalar theory, the nonlinear sigma model, and inverse string Kawai--Lewellen--Tye (KLT) functions. We derive the tree splitting formula from the equations of motion, showing that hidden zeros arise when the product of two currents solves the equations of motion, while corrections away from the zero reconstruct the full amplitude.  We further show that inverse KLT is the unique extension of $\mathrm{Tr}(\phi^3)$ with higher-point non-derivative interactions that preserves  hidden zeros.\end{abstract}

\maketitle
\pagestyle{plain}
\raggedbottom

\section{Introduction}\label{sec:introduction}

At a scattering pole, an intermediate particle goes on
shell and the residue factorizes into lower amplitudes.
Hidden zeros reveal structure away from poles: the
complete amplitude vanishes on special kinematic loci
through nontrivial cancellations between finite Feynman
diagrams~\cite{hidden}, occurring in a wide range of theories~\cite{zerosdoublecopy,exceptionalzeros,
expansionzeros,allstringzeros,openstring,integerrectangles,higherderivativezeros,massivezeros,
fermionzeros,formfactorstring,formfactorsplit,
cosmozeros,dualshuffle}.
They have been related to enhanced scaling under complex momentum shifts and used to reconstruct amplitudes without assuming locality or unitarity~\cite{zerosBCFW,BR,localityzeros}. This adds to a broader program in which amplitudes are uniquely determined by principles other than unitarity, with unitarity emerging as a consequence~\cite{gaugeuniqueness,Adleruniqueness,BCFWuniqueness,softuniqueness,UVbootstrap,CHYbootstrap}. These results suggest that hidden zeros may serve as fundamental organizing principles for quantum field theory.

Previous work has established near-zero, smooth, 
splitting relations, in which amplitudes reduce to products
of lower amplitudes or currents on restricted kinematic
loci~\cite{smooth,universal,universaldetailed,YMsplit,
YMsplitproof,diagrammatic,pionEOM,BCFWsplits,splitdiagrams,
onshellsplit,diagrammaticuniversal,higherderivativesplit,
allsplits,loopsplits,CEGMsplit,CEGMNLSM}.
A key precedent for the present work is the higher-order
splitting construction of Ref.~\cite{JP}, which proposed an identity valid at generic kinematics,
manifesting the skinny hidden zero.

In this work, we promote hidden zeros into a new recursive
construction of scattering amplitudes at generic kinematics, generalizing the above mentioned results.
Rather than merely constraining amplitudes on special loci,
the zeros organize the complete amplitude even when all
their defining conditions are relaxed. Remarkably, the
recursion takes a universal exact-splitting form,
schematically,
\begin{equation}
A^{\rm tree}=\sum_\alpha R_\alpha
\J_{\alpha,L}\J_{\alpha,R}\J_{\alpha,O},
\quad
\mathcal{I}^{\rm loop}=\sum_\alpha R_\alpha^{(1)}
A_\alpha \J_\alpha,
\end{equation}
where the prefactors $R_\alpha$ are precisely the tree or one-loop zero
conditions (or their trigonometric generalizations),
while the remaining factors are complete tree functions or amplitudes.
Their sum reconstructs the full amplitude, with every
term vanishing separately on the chosen zero locus.

For cubic scalars, pions, and cosmological wavefunction coefficients the tree-level exact splitting formula were already presented in Ref.~\cite{LR}, where they were used to generate new types of zero conditions. Here we extend the construction to  the inverse
string Kawai--Lewellen--Tye (KLT) functions, uncover their classical origin, and extend the recursion to one-loop
surface integrands in all three theories.
The resulting framework unifies previously known
restricted splitting relations and also naturally reproduces
the pion leading and subleading soft theorems.

We also propose a physical interpretation of hidden zeros and exact splitting, motivated by dual shuffle factorization, which separates diagrams along a zero line rather than across an on-shell propagator~\cite{dualshuffle} (see also~\cite{diagrammaticuniversal}). Surprisingly, on the hidden-zero locus, the product of two currents $\J_{\alpha,L}\J_{\alpha,R}$ satisfies the classical equation of motion. Away from this locus, the product generates a source proportional to the zero conditions $R_\alpha$. The response to this source is given by the third current $\J_{\alpha,O}$, reconstructing the full amplitude through the exact splitting formula.

Finally, we remark that the appearance of inverse KLT in this construction is not
accidental. Within matrix-scalar theories with an arbitrary
analytic kinetic function and non-derivative interactions, hidden zeros uniquely
select the one-parameter inverse KLT tree family, whose
undeformed limit is $\Tr(\phi^3)$ theory. Crucially, the
zeros determine not only the contact interactions but also
the trigonometric momentum dependence of the propagator
itself.\footnote{Related restrictions on KLT kernels follow from double-copy
consistency~\cite{KLTBootstrap,StringKLTBootstrap}.}

\section{Cubic scalars}\label{sec:cubic}

\subsection{Exact splitting}

The ordered cubic theory provides the basic example of the
splitting formula. The ordered $\Tr(\phi^3)$ tree amplitude
$A_n$, sums all planar cubic diagrams,
or equivalently all triangulations of the $n$-point
polygon. Each internal diagonal carries a propagator
$1/X_{ij}$, where
\begin{equation}
X_{ij}=(k_i+\cdots+k_{j-1})^2,
\qquad c_{ij}=-2k_i\cdot k_j.
\label{kinematics}
\end{equation}
All momenta are outgoing and massless, with $\sum_i k_i=0$. 
At four points,
\begin{equation}
A_4=\frac1{X_{13}}+\frac1{X_{24}}
=\frac{X_{13}+X_{24}}{X_{13}X_{24}}.
\label{cubic-four-main}
\end{equation}
The sum of the two exchange diagrams displays the hidden
zero $c_{13}=X_{13}+X_{24}=0$, away from poles.
Exact splitting extends this rearrangement to every
multiplicity and every rectangular zero.

Fix $3\leq j\leq n-1$. This choice selects the
 zero
\begin{equation}
c_{uv}=0,\qquad
1\leq u\leq j-2,\qquad j\leq v\leq n-1.
\end{equation}
The exact-splitting formula adapted to this zero
is~\cite{LR}
\begin{equation}
\boxed{
\begin{aligned}
A_n={}\sum_{a=1}^{j-2}\sum_{k=j+1}^{n}
R_{ak}
\times\J[P_L]\J[P_R]\J[P_O].
\end{aligned}}
\label{tree}
\end{equation}
Here $\J[P]$ is a complete lower amplitude with
its attaching propagator:
\begin{equation}
\J[P]=\frac{A[P]}{X_{p_1p_m}},
\qquad P=(p_1,\ldots,p_m),
\qquad \J[n,1]=1.
\label{current}
\end{equation}
The three lower polygons are
\begin{equation}
\begin{aligned}
P_L&=(a,\ldots,j),\qquad
P_R=(j-1,\ldots,k),\\
P_O&=(k,\ldots,n,1,\ldots,a).
\end{aligned}
\label{polygons}
\end{equation}
Each $A[P]$ is a complete on-shell lower-point amplitude, regarded
as a function of its internal channels. 

The prefactors are precisely sums of the defining zero conditions:
\begin{equation}
R_{ak}=X_{aj}+X_{j-1,k}-X_{ak}
=
\sum_{u=a}^{j-2}\sum_{v=j}^{k-1}c_{uv}.
\label{zeros}
\end{equation}
Each term vanishes on the chosen rectangle, while the splitting formula
holds at generic kinematics. If all cells except \(c_{1,n-1}\) are set
to zero, only \((a,k)=(1,n)\) contributes and we recover the near-zero split~\cite{hidden},
\begin{equation}
 A_n=\left(\frac{1}{X_{1j}}+\frac{1}{X_{j-1,n}}\right)
 A[1,\ldots,j]\,A[j-1,\ldots,n].
\end{equation}

For example, at 5-points, choose $j=3$ in Eq.~\eqref{tree}. The formula reads
\begin{equation}
 A_5=
 \frac{X_{13}+X_{25}}{X_{13}X_{25}}
 \left(\frac1{X_{24}}+\frac1{X_{35}}\right)
 +\frac{X_{13}+X_{24}-X_{14}}{X_{13}X_{24}X_{14}},
 \label{ex:cubic-tree-fivex}
\end{equation}
which expands to the five planar cubic trees. The two numerators in Eq.~\eqref{ex:cubic-tree-fivex} are
$c_{13}+c_{14}$ and $c_{13}$ so each 
term vanishes on the $j=3$ rectangle.

\subsection{Origin from the equation of motion}\label{sec:classical}
We now derive the splitting formula from the classical equation
$\Box\Phi=-\Phi^2$, motivated by dual shuffle
factorization, which naturally splits diagrams along a line joining $j-1$ and $n$, which also defines corresponding zero conditions~\cite{dualshuffle}. This suggests replacing
the shuffle by a product of the two associated classical currents,
$J_{ak}^{(0)}=J_{aj}J_{j-1,k}$. Surprisingly, this trial solution satisfies the equation of motion on the zero locus defined by legs $j-1$ and $n$. We will show that the correction
required by the equation of motion, away from the zero locus, instead generates precisely the
exact splitting formula of Eq.(\ref{tree}).

Here $J_{ab}=\J[a,\ldots,b]$ sums the ordered trees with external
legs $a,\ldots,b-1$, with $J_{a,a+1}=1$. Taking leg $n$ as the
root, the amplitude is obtained by removing the root propagator
and taking the root on shell
$A_n=\lim_{X_{1n}\to0}X_{1n}J_{1n}$. For each interval containing the marked leg $j-1$, write
\begin{equation}
J_{ak}=J_{ak}^{(0)}+\Delta_{ak}.
\label{classical-product-ansatz}
\end{equation}
To determine $\Delta_{ak}$, substitute Eq. \eqref{classical-product-ansatz}
into the ordered current equation
$X_{ab}J_{ab}=\sum_{a<d<b}J_{ad}J_{db}$. The terms that attach a
branch to either factor in the trial product sum to
$(X_{aj}+X_{j-1,k})J_{ak}^{(0)}$, while the inverse propagator of
the combined current contributes $X_{ak}J_{ak}^{(0)}$. Their
difference is the source for the correction:
\begin{equation}
\sigma_{ak}
=\bigl(X_{aj}+X_{j-1,k}-X_{ak}\bigr)J_{aj}J_{j-1,k}.
\label{classical-source-main}
\end{equation}
By Eq.~\eqref{zeros}, the prefactor is a sum of the $c$-variables
defining the zero.

At each cubic vertex only one sub-current contains the marked
leg, so the equation for $\Delta_{ak}$ is linear. Iterating it
attaches the remaining branches to each source in all planar ways.
After root amputation, their sum is the outside current $\J[P_O]$
on $P_O=(k,\ldots,n,1,\ldots,a)$, with $\J[n,1]=1$. Since the
trial product $J_{1j}J_{j-1,n}$ has no pole in $X_{1n}$, only the
correction contributes to the amplitude:
\begin{equation}
A_n=\sum_{a=1}^{j-2}\sum_{k=j+1}^{n}\sigma_{ak}\,\J[P_O].
\label{classical-amputation-main}
\end{equation}
This proves Eq.~\eqref{tree} at generic kinematics. On the
zero locus every source vanishes, so the trial product
already solves the current equation and $A_n=0$.

\section{The inverse string KLT kernel}\label{sec:klt}

The Kawai--Lewellen--Tye (KLT) relations express
closed-string tree amplitudes as sums of products of
ordered open-string amplitudes. Their kinematic weights
form the KLT kernel, whose matrix inverse is described
by scalar functions with two cyclic orderings~\cite{Mizera,intersection}.
We study matching orderings and denote the normalized
function by $\F_{\lambda,n}$, with $\F_{\lambda,3}=1$
and $\lambda=\pi\alpha'$.

These functions are generated at tree level by
\begin{equation}
\mathcal L_\lambda
=
\Tr\!\left[
\frac12\Phi
\frac{\tan[\lambda(-\Box)]}{\lambda}\Phi
-\sum_{h=0}^{\infty}
\frac{C_h\lambda^{2h}}{2h+3}\Phi^{2h+3}
\right],
\label{klt-lagrangian}
\end{equation}
where $C_h=\frac1{h+1}\binom{2h}{h}$ is a Catalan
number.
An internal line contributes
$\lambda\cot(\lambda X_{ij})$, and an ordered
$(2h+3)$-point vertex contributes $C_h\lambda^{2h}$~\cite{Mizera,surfaceKLT}.
The limit $\lambda\to0$ restores the ordinary
$\Tr(\phi^3)$ kinetic term and removes all higher
contacts.

\subsection{Exact splitting}

At four points, summing the two exchange diagrams gives
\begin{equation}
\begin{aligned}
\F_{\lambda,4}
={}&\lambda\cot(\lambda X_{13})
+\lambda\cot(\lambda X_{24})\\
={}&\frac{\lambda\sin[\lambda(X_{13}+X_{24})]}
{\sin(\lambda X_{13})\sin(\lambda X_{24})}.
\end{aligned}
\label{klt-four-main}
\end{equation}
The numerator vanishes at $X_{13}+X_{24}=m/\alpha'$,
with $m\in\mathbb Z$. The case $m=0$ recovers the familiar
field-theory zero, while nonzero $m$ gives additional zeros
at finite $\alpha'$.

For each $3\leq j\leq n-1$, the inverse KLT amplitude obeys
the following trigonometric counterpart of Eq.~\eqref{tree}:
\begin{equation}
\begin{aligned}
\F_{\lambda,n}={}&
\sum_{a=1}^{j-2}\sum_{k=j+1}^{n}
R_{\lambda;ak}
\times
\J_\lambda[P_L]\J_\lambda[P_R]\J_\lambda[P_O].
\end{aligned}
\label{finite}
\end{equation}
The lower polygons are those of Eq.~\eqref{polygons}.
For $P=(p_1,\ldots,p_m)$, define the attached functions
\begin{equation}
\J_\lambda[P]
=
\frac{\lambda\F_\lambda[P]}
{\sin(\lambda X_{p_1p_m})},
\qquad
\J_\lambda[n,1]=1.
\label{klt-current}
\end{equation}
Each $\F_\lambda[P]$ is the complete lower-point amplitude
generated by Eq.~\eqref{klt-lagrangian}, including all exchange
and contact contributions, with internal channels inherited
from the parent polygon. The attaching factors involve inverse
sines, while propagators within the lower amplitudes retain
their cotangent form.

The prefactor makes the zero condition explicit:
\begin{equation}
R_{\lambda;ak}
=\frac{\sin[\lambda(X_{aj}+X_{j-1,k}-X_{ak})]}{\lambda}.
\end{equation}
By Eq.~\eqref{zeros}, the argument of each sine is $\lambda$
times a sum of $c$-variables in the selected rectangle.
Thus imposing $\alpha'c_{uv}\in\mathbb Z$ throughout the
rectangle makes every term vanish~\cite{surfaceKLT,colored}.
Eq.~\eqref{finite} holds at generic kinematics, with all
zero conditions relaxed, and reduces term by term to the
cubic splitting formula as $\lambda\to0$.

\paragraph{Derivation from the equation of motion.}
To derive Eq.~\eqref{finite}, we use an equivalent field
variable whose classical equation is~\cite{intersection}
\begin{equation}
D_\tau(-\Box)\Phi=\frac{\Phi^2}{1-\tau\Phi},
\quad
D_\tau(X_{ij})=\frac{e^{\tau X_{ij}}-1}{\tau},
\label{classical-klt-eom-main}
\end{equation}
where $\tau=2i\lambda$. Appendix~\ref{appendix:klt-eom} gives the derivation
and the field redefinition relating this equation to the
Lagrangian in Eq.~\eqref{klt-lagrangian}.

\section{The nonlinear sigma model}\label{sec:nlsm}

\subsection{Exact splitting}
The NLSM describes massless Goldstone pions with derivative
interactions. Pure pion amplitudes vanish at odd multiplicity.

In the normalization $A_4^\pi=-(X_{13}+X_{24})$,
the ordered tree exact splitting identity is also given by the same expression in Eq.~(\ref{tree})~\cite{LR}
\begin{equation}
\begin{aligned}
A_n^\pi={}&\sum_{a=1}^{j-2}\sum_{k=j+1}^{n}
R_{ak}\times\J^\pi[P_L]\J^\pi[P_R]\J^\pi[P_O],
\end{aligned}
\label{pion}
\end{equation}
with $R_{ak}$ given by Eq.~(\ref{zeros}). The lower functions are
\begin{equation}
\J^\pi[P]=
\begin{cases}
A_m^\pi[P]/X_{p_1p_m},&m\text{ even},\\
(-1)^{p_1}A_m^{\rm ext}[P],&m\text{ odd}.
\end{cases}
\label{pionJ}
\end{equation}
The complete mixed amplitude $A_m^{\rm ext}$ has three
biadjoint scalars at $p_1,p_{m-1},p_m$, in that second
color-trace order, and pions elsewhere~\cite{extensions,LowYin2018}. We take $A_3^{\rm ext}=1$.

The pion amplitude was remarkably shown to follow from the cubic scalar amplitude through the 
$\delta$-shift relation~\cite{NLSM},
\begin{equation}
\begin{aligned}
&X_{ee}\to X_{ee}+\delta,\quad 
X_{oo}\to X_{oo}-\delta,\\
&A_n^\pi=\lim_{\delta\to\infty}
\delta^{n-2}A_n(X^{(\delta)}),
\end{aligned}
\label{pionlimit}
\end{equation}
where $e$ and $o$ denote even and odd labels, respectively,
and mixed-parity channels are unchanged. The same relation can be used to derive Eq.~(\ref{pion}) from Eq.~(\ref{tree}).

 The pion splitting formula Eq. \eqref{pion} also follows from the
classical field equation
\[
\Box\pi=\{C,\Box\pi\}-\{\pi,\Box C\},
\qquad C=\tfrac12(1-\sqrt{1-4\pi^2}),
\]
where the composite field $C$ collects the pion derivative interactions. Appendix~\ref{appendix:examples} shows examples, while Appendix~\ref{appendix:pion-eom} proves the required projection, and also directly derives it from the equation of motion.

Eq.~\eqref{pion} advances the exact-factorization program
of Ref.~\cite{pionEOM} by providing an explicit identity for
the full pion tree amplitude at arbitrary multiplicity and
generic kinematics.

\subsection{Soft limits}
The exact splitting formula in Eq.\eqref{pion} directly yields both the
single- and double-soft theorems for pions.
In the single-soft limit, the zero prefactors make the
Adler zero manifest, while the current sums combine into
the mixed amplitudes of the subleading theorem~\cite{extensions}.
 Schematically,
\begin{equation}
A_n^\pi(\epsilon p,\ldots)
=
\epsilon\sum_i 2p\cdot k_i\,
\underbrace{
\left.\sum_{k>i}\J^\pi\J^\pi\right|_{\epsilon=0}
}_{A_{n-1}^{\rm ext}}
+\mO(\epsilon^2).
\end{equation}
More details are given in Appendix~\ref{appendix:soft}. In the adjacent double-soft limit, only two terms survive, and one can show they reduce to the hard pion amplitude times a soft factor, giving the leading double-soft theorem.

\section{One-loop surface integrands}\label{sec:one-loop}
In this section we extend the splitting formulas to one-loop surface integrands. 
In this case, we find that exact splitting expresses the loop integrand in terms of trees. 

\subsection{Cubic scalars}

For cubic scalars, a one-loop surface integrand is a sum over
triangulations of a polygon with one puncture $z$, each dual
to a planar one-loop Feynman diagram~\cite{allsplits,BR}.
Each internal curve represents a propagator, and the weight
of a triangulation is the product of these propagators.
A curve joining boundary vertices $i$ and $j$ carries
$1/X_{ij}$. The curves associated with $X_{ij}$ and $X_{ji}$
pass on opposite sides of the puncture and are assigned
independent variables.

We also assign the puncture an auxiliary parity $\sigma=\pm$.
A curve from vertex $i$ to the puncture carries $1/Y_i^{\sigma}$,
with the same parity used throughout each triangulation
[Fig.~\ref{fig:loop-geometrya}]. Summing both sectors gives
\begin{equation}
\I_n^{(1)}=\I_n^+(Y^+)+\I_n^-(Y^-).
\end{equation}

The usual momentum-space integrand is obtained by first
discarding the tadpole and external-bubble contributions
associated with $X_{ii}$ and $X_{i+1,i}$, whose loop integrals
are scaleless in dimensional regularization. We then identify
the remaining curve variables with momentum invariants $Y_i^\pm=(\ell+k_1+\cdots+k_{i-1})^2$
where $\ell$ is the
loop momentum.

For example, at two points, summing the three
triangulations in Fig.~\ref{fig:loop-geometrya} over
$\sigma=\pm$ gives

\begin{figure}[t]
\centering
\begin{tikzpicture}[x=.62cm,y=.62cm,
  every node/.style={font=\scriptsize},line width=.45pt]
\foreach \xx in {0,4,8}{
 \draw (\xx,0) circle(1.25);
 \fill (\xx-1.25,0) circle(.04);
 \node[left,inner sep=1.5pt] at (\xx-1.25,0){$1$};
 \fill (\xx+1.25,0) circle(.04);
 \node[right,inner sep=1.5pt] at (\xx+1.25,0){$2$};
 \fill (\xx,0) circle(.045);
 \node[red,above,inner sep=1.5pt] at (\xx,.04){$z$};
}
\draw[cyan!70!blue,thick] (-1.25,0)--(0,0)--(1.25,0);
\node[cyan!70!blue,below,inner sep=2pt] at (-.64,0){$Y_1^\sigma$};
\node[cyan!70!blue,below,inner sep=2pt] at (.64,0){$Y_2^\sigma$};
\node at (0,-1.97){$\displaystyle\frac1{Y_1^\sigma Y_2^\sigma}$};
\draw[cyan!70!blue,thick] (2.75,0)--(4,0);
\node[cyan!70!blue,below,inner sep=2pt] at (3.38,0){$Y_1^\sigma$};
\draw[gray] (2.75,0) .. controls (3.2,1.13) and (4.85,1.02) .. (4.85,0)
 .. controls (4.85,-1.02) and (3.2,-1.13) .. (2.75,0);
\node[fill=white,inner sep=.6pt] at (4.12,.84){$X_{11}$};
\node at (4,-1.97){$\displaystyle\frac1{X_{11}Y_1^\sigma}$};
\draw[cyan!70!blue,thick] (8,0)--(9.25,0);
\node[cyan!70!blue,below,inner sep=2pt] at (8.62,0){$Y_2^\sigma$};
\draw[gray] (9.25,0) .. controls (8.8,1.13) and (7.15,1.02) .. (7.15,0)
 .. controls (7.15,-1.02) and (8.8,-1.13) .. (9.25,0);
\node[fill=white,inner sep=.6pt] at (7.88,.84){$X_{22}$};
\node at (8,-1.97){$\displaystyle\frac1{X_{22}Y_2^\sigma}$};
\end{tikzpicture}
\caption{The three two-point triangulations in either
puncture-parity sector $\sigma=\pm$; summing both sectors gives
Eq.~\eqref{loop-two-main}}
\label{fig:loop-geometrya}
\end{figure}
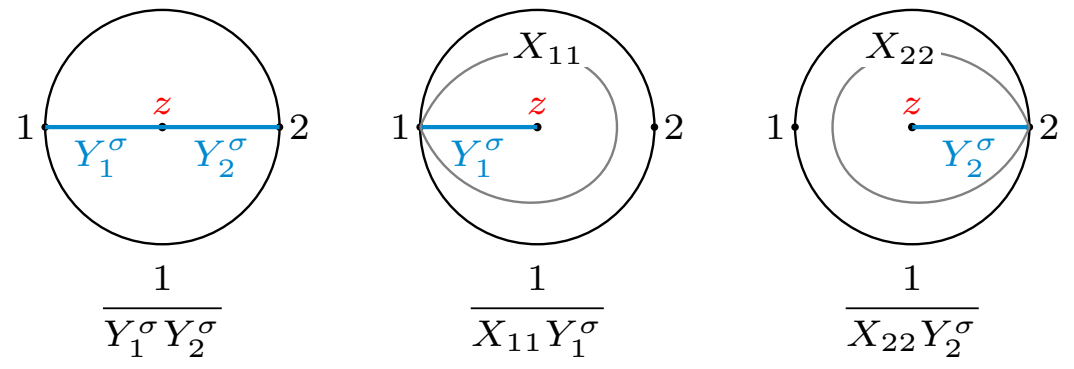
\begin{equation}
\I_2^{(1)}=\sum_{\sigma=\pm}\left(
\frac1{Y_1^\sigma Y_2^\sigma}
+\frac1{X_{11}Y_1^\sigma}
+\frac1{X_{22}Y_2^\sigma}\right).
\label{loop-two-main}
\end{equation}

We find the full one-loop surface integrand admits an exact splitting into
complete trees at generic kinematics. For arbitrary multiplicity,
\begin{equation}
\boxed{
\I_n^{(1)}
=\sum_{1\leq j\leq i\leq n}
R^{(1)}_{ij}\,
\frac{A[P_T]}{Y_j^-Y_i^+}\,\J[P_O]
}
\label{loop}
\end{equation}
where $R^{(1)}_{ij}=Y_i^++Y_j^--X_{ij}$. The inner and outside
polygons are $P_T=(L,j,\ldots,i,R)$, and 
$P_O=(i,\ldots,n,1,\ldots,j)$. As at tree level, $\J[P_O]=A[P_O]/X_{ij}$ includes its attaching
propagator, with $\J[n,1]=1$. Examples are given in Appendix~\ref{appendix:examples}.

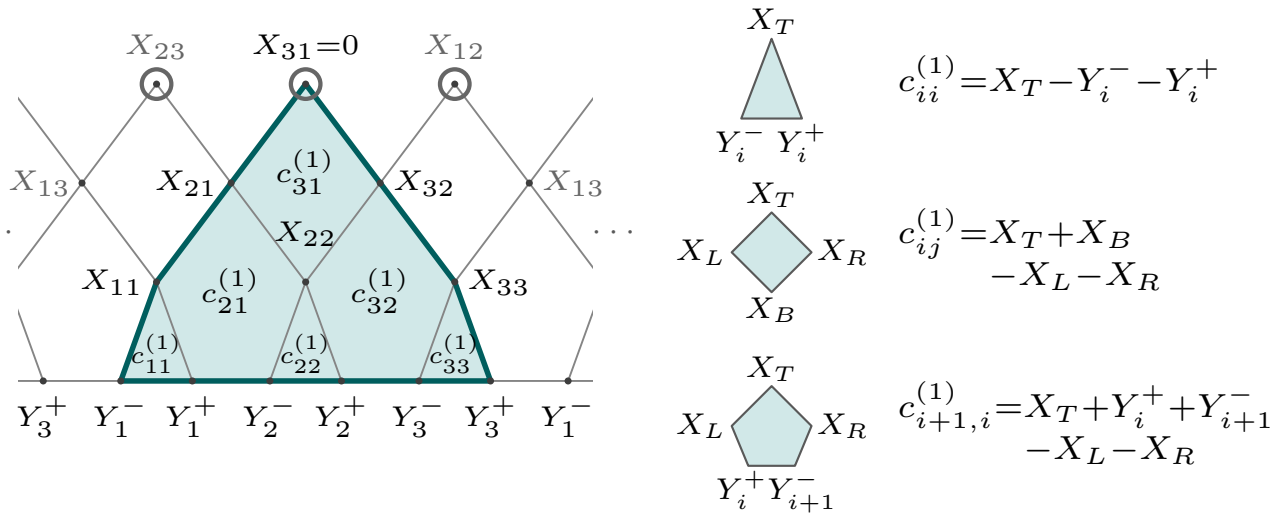
\begin{figure}[b]
\centering
\resizebox{.999\columnwidth}{!}{%
\begin{tikzpicture}[x=.55cm,y=.73cm,line cap=round,line join=round,
  every node/.style={font=\scriptsize,inner sep=0pt},
  mesh/.style={draw=black!48,line width=.4pt},
  facet fill/.style={fill=teal!18},
  mountain/.style={draw=teal!74!black,line width=1.15pt},
  vertex label/.style={right=2.5pt},
  outer label/.style={text=black!60}]
  \begin{scope}[every node/.append style={font=\fontsize{6.4}{7.5}\selectfont}]
  \fill[facet fill] (-.48,0)--(.48,0)--(0,1)--cycle;
  \fill[facet fill] (1.52,0)--(2.48,0)--(2,1)--cycle;
  \fill[facet fill] (3.52,0)--(4.48,0)--(4,1)--cycle;
  \fill[facet fill] (.48,0)--(1.52,0)--(2,1)--(1,2)--(0,1)--cycle;
  \fill[facet fill] (2.48,0)--(3.52,0)--(4,1)--(3,2)--(2,1)--cycle;
  \fill[facet fill] (2,1)--(3,2)--(2,3)--(1,2)--cycle;

  \begin{scope}
    \clip (-1.85,-.09) rectangle (5.85,3.09);
    \draw[mesh] (-4,0)--(8,0);
    \foreach \x in {-6,-4,...,10}{
      \draw[mesh] (\x-.48,0)--(\x,1)--(\x+2,3);
      \draw[mesh] (\x+.48,0)--(\x,1)--(\x-2,3);
    }
  \end{scope}

  \draw[mountain] (-.48,0)--(0,1)--(2,3)--(4,1)--(4.48,0)--cycle;

  \foreach \h/\xs in {1/{0,2,4},
                       2/{-1,1,3,5},3/{0,2,4}}
    \foreach \x in \xs
      \fill[black!76] (\x,\h) circle[radius=.74pt];
  \foreach \x in {0,2,4}
    \draw[black!60,line width=.9pt] (\x,3) circle[radius=3.1pt];

  \foreach \x/\ij in {0/11,2/22,4/33}
    \node[scale=.86] at (\x,.28) {$c^{(1)}_{\ij}$};
  \foreach \x/\h/\ij in {1/.91/21,3/.91/32,2/2.12/31}
    \node at (\x,\h) {$c^{(1)}_{\ij}$};
  \foreach \x/\h/\ij in {0/1/11,1/2/21}
    \node[left=2.5pt] at (\x,\h) {$X_{\ij}$};
  \foreach \x/\h/\ij in {4/1/33,3/2/32}
    \node[vertex label] at (\x,\h) {$X_{\ij}$};
  \node at (2,1.49) {$X_{22}$};
  \node[above=5pt] at (2,3) {$X_{31}{=}0$};

  \node[vertex label,outer label] at (5,2) {$X_{13}$};
  \node[left=2pt,outer label] at (-1,2) {$X_{13}$};
  \foreach \x/\ij in {0/23,4/12}
    \node[above=5pt,outer label] at (\x,3) {$X_{\ij}$};

  \foreach \x/\i/\sign in {-1.52/3/+,
    -.48/1/-,.48/1/+,1.52/2/-,2.48/2/+,3.52/3/-,
    4.48/3/+,5.52/1/-}{
    \fill[black!76] (\x,0) circle[radius=.74pt];
    \node[below=3pt] at (\x,0) {$Y_{\i}^{\sign}$};
  }
  \foreach \x in {-2.17,6.17}
    \node[text=black!65] at (\x,1.5) {$\cdots$};

  \end{scope}

  \begin{scope}[shift={(8.25,1.30)},x=.40cm,y=.42cm,
    facet/.style={draw=black!65,line width=.45pt,scale=.8,
      every node/.append style={scale=.9}},
    formula/.style={anchor=west,inner sep=0pt},
    every node/.append style={inner sep=1pt}]
   
    \begin{scope}[shift={(0,3.05)}]
      \draw[facet,facet fill]
        (0,.875) node[above] {$X_T$}
        --(-.70,-.875) node[below] {$Y_i^-$}
        --(.70,-.875) node[below] {$Y_i^+$} --cycle;
      \node[formula] at (2.31,0)
        {$c^{(1)}_{ii}{=}X_T{-}Y_i^{-}{-}Y_i^{+}$};
    \end{scope}

    \draw[facet,facet fill]
      (0,.875) node[above] {$X_T$}
      --(-.92,0) node[left] {$X_L$}
      --(0,-.875) node[below] {$X_B$}
      --(.92,0) node[right] {$X_R$} --cycle;
    \node[formula] at (2.31,0) {$\begin{array}{@{}r@{}l@{}}
      c^{(1)}_{ij}{=}&X_T{+}X_B\\
      &{-}X_L{-}X_R
    \end{array}$};

    \begin{scope}[shift={(0,-3.05)}]
      \draw[facet,facet fill]
        (0,.875) node[above] {$X_T$}
        --(-.92,0) node[left] {$X_L$}
        --(-.538,-.875) node[below,xshift=-1.5pt] {$Y_i^+$}
        --(.538,-.875) node[below,xshift=1.7pt] {$Y_{i+1}^-$}
        --(.92,0) node[right] {$X_R$} --cycle;
      \node[formula] at (2.31,0) {$\begin{array}{@{}r@{}l@{}}
        c^{(1)}_{i+1,i}{=}&X_T{+}Y_i^{+}{+}Y_{i+1}^{-}\\
        &{-}X_L{-}X_R
      \end{array}$};
    \end{scope}
  \end{scope}
\end{tikzpicture}%
}%
\caption{ A three-point mountain on the
kinematic mesh~\cite{BR}. Setting every shaded cell to zero gives the
mountain zero for the integrand.}
\label{fig:loop-geometryb}
\end{figure}

As in Eq.~\eqref{tree}, each prefactor is a sum of local zero
conditions. On the loop mesh, the cells $c^{(1)}_{ab}$ satisfy
\begin{equation}
R^{(1)}_{ij}
=Y_i^++Y_j^--X_{ij}
=-\sum_{j\leq b\leq a\leq i}c^{(1)}_{ab}.
\label{loop-zero-conditions}
\end{equation}
The cells form a triangular region called a mountain~\cite{BR},
shown at three points in Fig.~\ref{fig:loop-geometryb}.
Setting every cell of the mountain to zero makes
Eq.~\eqref{loop} vanish term by term. Allowing only the peak cell
to be nonzero recovers the restricted one-loop split of
Ref.~\cite{BR}. 

\emph{Tree splitting from a loop residue.}
The residue at $Y_r^+=0$, with $1\leq r<n$, opens the loop into
an $(n+2)$-point tree and takes the mountain zero to a rectangular
tree zero~\cite{BR}. For $j\leq r<i$, the same cut is the diagonal
$(r,R)$ of $P_T$, and therefore splits the inner tree into two trees. Together
with $\J[P_O]$, these give the three factors of the tree splitting
formula. Varying $r$ recovers every splitting position in
Eq.~\eqref{tree}. Conversely, agreement on all radial residues fixes
the cubic integrand, since every cubic diagram contains a radial
propagator. The proof is given in Appendix~\ref{app:loop-cut-proof}.

\subsection{Inverse KLT}

For inverse KLT, the one-loop surface integrand is the sum of the
two puncture-parity sectors,
$\I_{n,\lambda}^{(1)}
=\I_{n,\lambda}^+(Y^+)+\I_{n,\lambda}^-(Y^-)$.
Each sector uses the tree vertices and propagators of
Eq.~\eqref{klt-lagrangian}, together with the one-loop contact
interactions prescribed in Ref.~\cite{surfaceKLT}.

The exact splitting formula has the same form as in the cubic theory
\begin{equation}
\I_{n,\lambda}^{(1)}
=\sum_{1\leq j\leq i\leq n}
R^{(1)}_{\lambda;ij}\,
\frac{\lambda^2\F_\lambda[P_T]}
     {\sin(\lambda Y_j^-)\sin(\lambda Y_i^+)}\,
\J_\lambda[P_O]
\label{loop-sine}
\end{equation}
with
\begin{equation}
R^{(1)}_{\lambda;ij}
=\frac{\sin\!\left[\lambda(Y_i^++Y_j^--X_{ij})\right]}{\lambda}.
\label{loop-sine-trees}
\end{equation}
As at tree level, each attaching channel $X$ contributes
$\lambda/\sin(\lambda X)$. The outside current
$\J_\lambda[P_O]=\lambda\F_\lambda[P_O]/\sin(\lambda X_{ij})$
carries the boundary attachment $X_{ij}$. At the peak,
$\J_\lambda[n,1]=1$.

Away from poles, every term vanishes on the locus
$\lambda(Y_i^++Y_j^--X_{ij})\in\pi\mathbb Z$ for all
$1\leq j\leq i\leq n$. As $\lambda\to0$, each attaching factor
reduces to its cubic propagator and $R^{(1)}_{\lambda;ij}$ reduces
to $R^{(1)}_{ij}$, recovering Eq.~\eqref{loop} term by term.

At finite $\lambda$, the radial cuts fix all terms containing radial
propagators. To isolate the remaining terms, set
$\lambda Y_i^\pm=\pi/2$ for all $i$. Then the radial propagators
$\lambda\cot(\lambda Y_i^\pm)$ vanish, while the boundary channels
$X_{ij}$ remain arbitrary. On this locus, Eq.~\eqref{loop-sine}
reproduces the one-loop contact interactions of
Ref.~\cite{surfaceKLT} with all their tree attachments
(Appendix~\ref{app:loop-cut-proof}).
\subsection{The nonlinear sigma model}

For even $n$, apply the shifts in Eq.~\eqref{pionlimit} to the
boundary channels and shift the radial channels as
\begin{equation}
(Y_i^\pm)^{(\delta)}
=Y_i^\pm+\frac{\delta}{2}\bigl[(-1)^i\mp1\bigr].
\label{pion-radial-shifts}
\end{equation}
Writing $\I_n^{(1)}(\delta)$ for the cubic surface integrand at the
shifted arguments, we obtain the pion surface integrand as
\begin{equation}
\I_{n,\pi}^{(1)}
=\lim_{\delta\to\infty}\delta^n\I_n^{(1)}(\delta).
\end{equation}
The shifts preserve each mountain numerator
$R^{(1)}_{ij}=Y_i^++Y_j^--X_{ij}$. The cubic splitting formula Eq. \eqref{loop} therefore gives
\begin{equation}
\I_{n,\pi}^{(1)}
=\sum_{1\leq j\leq i\leq n}
R^{(1)}_{ij}\,
\left[\frac{A[P_T]}{Y_j^-Y_i^+}\right]_{\pi}
\left[\J[P_O]\right]_{\pi}
.
\label{pion-loop}
\end{equation}
For either attached tree factor $F[P]$, the pion projection is
\begin{equation}
\bigl[F[P]\bigr]_{\pi}
=\lim_{\delta\to\infty}\delta^{|P|-2}F^{(\delta)}[P],
\label{pion-projected-factors}
\end{equation}
where $|P|$ counts the vertices of $P$. The inherited shifts act
on the complete tree and its attaching propagators. 

Appendix~\ref{app:loop-cut-proof} proves
Eq.~\eqref{pion-loop}, while Appendix~\ref{appendix:pion-loop}
defines the projected tree factors, including their
composite insertions and contact terms.

\section{Uniqueness of inverse KLT from hidden zeros}\label{sec:selection}
In this section we show that the inverse string KLT theory is uniquely selected by zeros from a general class. Consider one massless
matrix scalar with single-trace, momentum-independent
interactions and the Lagrangian
\begin{equation}
\mathcal L
=
\Tr\!\left[
\frac12\Phi D(-\Box)\Phi
-\sum_{m=3}^{\infty}\frac{g_m}{m}\Phi^m
\right].
\label{selection-lagrangian}
\end{equation}
The constants $g_m$ are the ordered vertex weights, and
an internal line carrying channel $X_{ij}$ contributes
$1/D(X_{ij})$. The kinetic function $D$ is analytic near
the origin, with $D(0)=0$ and $D'(0)=1$, so its
propagator has a massless pole of unit residue,
$1/D(X_{ij})=1/X_{ij}+O(1)$.
Ordinary cubic theory has $D(X_{ij})=X_{ij}$ and
$g_m=0$ for $m>3$. At every multiplicity $n\geq4$, we require the amplitude
to vanish at regular points of every  zero locus. We normalize  $g_3=1$. At four points, 
\begin{equation}
\A_4
=
\frac1{D(X_{13})}
+\frac1{D(X_{24})}
+g_4.
\label{selection-four}
\end{equation}
Its vanishing at $X_{24}=-X_{13}$ fixes the even
part of the propagator. In particular, the function
\begin{equation}
P(X_{ij})=\frac1{D(X_{ij})}+\frac{g_4}{2}
\label{selection-P}
\end{equation}
is odd and has the expansion
$P(X_{ij})=1/X_{ij}+O(X_{ij})$.
Here $P$ denotes the propagator with its constant
part removed.

The five-point amplitude contains five cubic exchange
diagrams, and
on the rectangle $c_{13}=c_{14}=0$, vanishing requires
\begin{equation}
P(X_{13}+X_{24})
\bigl[P(X_{13})+P(X_{24})\bigr]=P(X_{13})P(X_{24})-\kappa,
\label{selection-addition}
\end{equation}
where $\kappa=g_5-\frac54g_4^2$. This addition law
determines the remaining momentum dependence.
Expanding it about $X_{24}=0$, using
the unit-residue pole, gives
$P'(X_{ij})+P(X_{ij})^2+\kappa=0$. 
To solve the equation, write $P=S'/S$.
The equation becomes $S''+\kappa S=0$, which gives
$S(X_{ij})=\sin(\sqrt{\kappa}X_{ij})/\sqrt{\kappa}$.
Consequently,
\begin{equation}
\frac1{D(X_{ij})}
=
\sqrt{\kappa}\cot(\sqrt{\kappa}X_{ij})
-\frac{g_4}{2}.
\label{selection-main}
\end{equation}
The four- and five-point zeros therefore fix the momentum
dependence up to $\kappa$ and a constant propagator shift, while the higher-point zeros fix the remaining interactions. The resulting tree amplitudes is in fact independent of $g_4$ at fixed $\kappa$. Thus  zeros select the one-parameter,
matching-order inverse KLT tree family within the class of Lagrangians in Eq.~\eqref{selection-lagrangian}. The complete interaction tower and its uniqueness are
derived in Appendix~\ref{appendix:interaction-tower}.

\section{Outlook}\label{sec:outlook}

We have introduced exact splitting, a recursion that reconstructs amplitudes at generic kinematics while making the chosen hidden zero manifest in every term. The classical equations of motion explain both why the zero conditions appear as prefactors, and the product of three currents.  The product of two currents generates a source, and the third sums all tree attachments to it. On the zero locus, the product of two currents satisfies the equation of motion.

At one loop, exact splitting expresses surface integrands entirely in terms of trees and reproduces the tree identities on radial cuts. Whether any structure survives after the loop momentum is integrated remains open. A further question is whether the hidden zeros of Yang--Mills amplitudes admit a similar recursion.

Such recursive formulas also reveal zeros beyond the rectangular loci~\cite{LR}. The same procedure can now be used to generate potentially new zeros even at one loop.
\medskip
\paragraph{Acknowledgments}
YL is supported in part by a Technion fellowship. LR is supported by the National Natural Science Foundation
of China General Program No. 12475070 and the Beijing
Natural Science Foundation International Scientist Project
No. IS24014. AI tools were used as research and coding
assistants for exploratory calculations, checks, and
manuscript editing. The authors are responsible for the
scientific content.

\bibliographystyle{apsrev4-2}
\bibliography{references}

\makeatletter
\immediate\write\@auxout{\string\citation{refsTitleControl}}
\makeatother

\clearpage

\onecolumngrid
\appendix

\section{Low-point examples}
\label{appendix:examples}

\subsection{Cubic scalars}

\paragraph{Two points at one loop.}
We retain independent $Y_i^+$ and $Y_i^-$, as well as the
tadpole channels $X_{ii}$. Since $X_{21}=0$, Eq.~\eqref{loop}
has three terms:
\begin{equation}
 \I_2^{(1)}={} \frac{Y_1^++Y_1^--X_{11}}{Y_1^+Y_1^-X_{11}} +\frac{Y_2^++Y_2^--X_{22}}{Y_2^+Y_2^-X_{22}} +\frac{Y_2^++Y_1^-}{Y_2^+Y_1^-} \left(\frac1{Y_1^+}+\frac1{Y_2^-}\right).
 \label{ex:cubic-loop-two}
\end{equation}
The last term contains $1/(Y_1^+Y_1^-)$ and
$1/(Y_2^+Y_2^-)$, which cancel the negative terms from the
first two numerators. The result is
\begin{equation}
 \I_{2}^{(1)}=
 \sum_{\sigma=\pm}\left[
 \frac1{Y_1^\sigma Y_2^\sigma}
 +\frac1{X_{11}Y_1^\sigma}
 +\frac1{X_{22}Y_2^\sigma}\right].
 \label{ex:cubic-loop-two-expanded}
\end{equation}
Thus each sector contains one bubble and two tadpoles.

\subsection{Inverse KLT}

\paragraph{Two points at one loop.}
The three terms in Eq.~\eqref{loop-sine} are
\begin{equation}
\begin{aligned}
 \I_{2,\lambda}^{(1)}=
 \frac{\lambda^2\sin[\lambda(Y_1^++Y_1^--X_{11})]}
      {\sin(\lambda Y_1^+)\sin(\lambda Y_1^-)\sin(\lambda X_{11})}
 +\frac{\lambda^2\sin[\lambda(Y_2^++Y_2^--X_{22})]}
      {\sin(\lambda Y_2^+)\sin(\lambda Y_2^-)\sin(\lambda X_{22})}
 +\frac{\lambda^2\sin[\lambda(Y_2^++Y_1^-)]}
      {\sin(\lambda Y_2^+)\sin(\lambda Y_1^-)}
 [\cot(\lambda Y_1^+)+\cot(\lambda Y_2^-)].
\end{aligned}
\label{ex:klt-loop-two}
\end{equation}
After cancellation of the mixed-sector products, this becomes
\begin{equation}
 \I_{2,\lambda}^{(1)}=\lambda^2\sum_{\sigma=\pm}
 \big[\cot(\lambda Y_1^\sigma)\cot(\lambda Y_2^\sigma)
 +\cot(\lambda X_{11})\cot(\lambda Y_1^\sigma)
 +\cot(\lambda X_{22})\cot(\lambda Y_2^\sigma)+1\big].
 \label{ex:klt-loop-two-expanded}
\end{equation}
Each sector contains the bubble, two tadpoles, and the
loop-order contact $\lambda^2$ fixed by the surface prescription.

\subsection{Pions}

\paragraph{Five and six points.}
The pure-pion five-point amplitude vanishes, $A_5^\pi=0$.
The nonzero five-point lower function in six-pion splitting
instead has three biadjoint scalars at $2,5,6$:
\begin{equation}
 A_5^{\rm ext}[2,3,4,5,6]
 =1-\frac{X_{24}+X_{35}}{X_{25}}
   -\frac{X_{35}+X_{46}}{X_{36}}.
 \label{ex:pion-mixed-five}
\end{equation}
The constant and the two poles are its contact and exchange
terms. Substituting this function and the four-pion amplitudes
into Eq.~\eqref{pion}, with $j=3$, gives
\begin{equation}
\begin{aligned}
 A_6^\pi=-(X_{13}+X_{26})
 \left[1-\frac{X_{24}+X_{35}}{X_{25}}
          -\frac{X_{35}+X_{46}}{X_{36}}\right]
 +\frac{X_{13}+X_{24}-X_{14}}{X_{14}}(X_{46}+X_{15})
 -\frac{X_{13}+X_{25}-X_{15}}{X_{25}}(X_{24}+X_{35}).
\end{aligned}
\label{ex:pion-six}
\end{equation}
The source numerators are $c_{13}+c_{14}+c_{15}$,
$c_{13}$, and $c_{13}+c_{14}$, so the selected zero is
manifest in each term. Expanding the shifted fourteen
cubic trees and extracting the coefficient of $\delta^{-4}$
gives the same expression.

\paragraph{Two points at one loop.}
At $n=2$, Eq.~\eqref{pion-radial-shifts} sends
$Y_1^+\to Y_1^+-\delta$, $Y_2^-\to Y_2^-+\delta$,
$X_{11}\to X_{11}-\delta$, and $X_{22}\to X_{22}+\delta$;
$Y_2^+,Y_1^-$ are unchanged. Applying this projection to the
three cubic terms in Eq.~\eqref{ex:cubic-loop-two} gives
\begin{equation}
 \mathcal{I}_{2,\pi}^{(1)}={}\frac{Y_1^++Y_1^--X_{11}}{Y_1^-} +\frac{Y_2^++Y_2^--X_{22}}{Y_2^+} -\frac{(Y_2^++Y_1^-)(Y_1^++Y_2^-)}{Y_2^+Y_1^-} ={}2-\frac{Y_1^++X_{22}}{Y_2^+} -\frac{Y_2^-+X_{11}}{Y_1^-}.
 \label{ex:pion-loop-two}
\end{equation}
The first expression retains the mountain numerators.

\section{Classical derivation of exact splitting}
\label{appendix:classical}

\subsection{Cubic scalars}
\label{appendix:cubic-eom}

We first derive the current equation and then identify the
contribution of each source to the amplitude. This also fixes
the conventions used for the other theories.
For the matrix equation $-\Box\Phi=\Phi^2$, one can perturbatively solve it by expanding over the on-shell plane waves $E_i=T_i e^{ik_i\cdot x}$:
\begin{equation}
 \Phi(x)=\sum_i E_i+\sum_{i,j}J_{ij}^{\rm word}E_iE_j+\cdots .
 \label{ceom:perturbiner}
\end{equation}
The matrices retain the color ordering. Let $J_{ab}$ denote the coefficient of $E_a\cdots E_{b-1}$;
in particular, $J_{a,a+1}=1$. Acting with $-\Box$ multiplies
this coefficient by $X_{ab}$. Substituting Eq. \eqref{ceom:perturbiner} into the field equation and equating the coefficients gives:
\begin{equation}
 X_{ab}J_{ab}=\sum_{d=a+1}^{b-1}J_{ad}J_{db}.
 \label{ceom:bg}
\end{equation}
These coefficients include the source propagator. The amplitude
is $A_n=\lim_{X_{1n}\to0}X_{1n}J_{1n}$.

Select external particle $h=j-1$. The terms in Eq.~\eqref{ceom:bg}
separate according to which factor contains $h$:
\begin{equation}
 X_{ak}J_{ak}
 =\sum_{b=a+1}^{h}J_{ab}J_{bk}
  +\sum_{b=j}^{k-1}J_{ab}J_{bk}.
 \label{ceom:marked}
\end{equation}
Only one factor in each term contains the plane wave.
We test $J^{(0)}_{ak}=J_{aj}J_{hk}$ and write
$J_{ak}=J^{(0)}_{ak}+\Delta_{ak}$. The two factors satisfy
\begin{equation}
 X_{aj}J_{aj}=\sum_{b=a+1}^{h}J_{ab}J_{bj},\qquad X_{hk}J_{hk}=\sum_{b=j}^{k-1}J_{hb}J_{bk}.
 \label{ceom:sides}
\end{equation}
Substituting the terms independent of $\Delta_{ak}$ sum, by Eq. \eqref{ceom:sides}, to $(X_{aj}+X_{hk})J^{(0)}_{ak}$. Subtracting its
contribution $X_{ak}J^{(0)}_{ak}$ from the left gives
\begin{equation}
 X_{ak}\Delta_{ak}={}\sigma_{ak} +\sum_{b=a+1}^{h}J_{ab}\Delta_{bk} +\sum_{b=j}^{k-1}\Delta_{ab}J_{bk},\qquad \sigma_{ak}={}(X_{aj}+X_{hk}-X_{ak})J_{aj}J_{hk}.
 \label{ceom:response}
\end{equation}
The boundary values are $\Delta_{aj}=\Delta_{hk}=0$.
Eq.~\eqref{ceom:response} determines the correction
recursively from currents with fewer external legs; no nonlinear terms in
$\Delta$ occur.

To identify the third factor, follow the contribution of one
proper source $\sigma_{ak}$, with $(a,k)\ne(1,n)$. It first
enters as $\sigma_{ak}/X_{ak}$. Every subsequent term of
Eq.~\eqref{ceom:response} attaches complete tree currents except leg $h$. Contracting the source sub-diagram
by an external leg $(a,k)$ leaves the outside current on
$(1,\ldots,a,k,\ldots,n)$, with the source propagator $1/X_{ak}$ kept explicit. Therefore
\begin{equation}
 \Delta_{1n}^{[a,k]} =\frac{\sigma_{ak}}{X_{ak}} \J[1,\ldots,a,k,\ldots,n] =\frac{\sigma_{ak}A[P_O]}{X_{ak}X_{1n}}.
 \label{ceom:outside}
\end{equation}
The second equality uses the cyclic ordering
$P_O=(k,\ldots,n,1,\ldots,a)$. Each outside tree occurs once:
the path from its source edge to the root specifies the
successive attachments. The full-interval source instead
contributes $\sigma_{1n}/X_{1n}$ directly.

The trial product has no $X_{1n}$ pole. Amputating the
correction therefore gives
\begin{equation}
 A_n=\sum_{a=1}^{j-2}\sum_{k=j+1}^{n}
       \sigma_{ak}\J[P_O],\qquad \J[n,1]=1,
 \label{ceom:splitting}
\end{equation}
which is Eq.~\eqref{tree}. The unit term involves no division
by $X_{1n}=0$. On a regular rectangular zero every source
vanishes, so the unique correction with the stated boundary
values is zero.

Eqs. \eqref{ceom:perturbiner} and \eqref{ceom:bg} give the standard perturbiner construction of scalar Berends–Giele currents \cite{Berends:1987me,MizeraSkrzypekPerturbiner}. Substituting the trial ansatz \(J_{aj}J_{hk}\) produces the source that is proportional to the rectangular-zero condition. Summing its tree attachments and amputating the root gives the three-current splitting formula Eq. \eqref{ceom:splitting}.

\subsection{Inverse KLT}
\label{appendix:klt-eom}

Only the kinetic function and the marked interaction change.
Consider the generating equation
\begin{equation}
 D_\tau(-\Box)\Phi=V(\Phi),\qquad V(\Phi)=\frac{\Phi^2}{1-\tau\Phi},\qquad D_\tau(X)=\frac{e^{\tau X}-1}{\tau}.
 \label{keom:equation}
\end{equation}
Here $D_\tau(0)=0$, $D_\tau'(0)=1$, and the ordered
$(r+1)$-point vertex is $\tau^{r-2}$. Expanding in external legs gives the Berends–Giele currents \(J_{ab}^{(\tau)}\), normalized by \(J_{a,a+1}^{(\tau)}=1\).

Let $\Phi$ denote the classical solution generated by unmarked external legs and set
$U=\Phi/(1-\tau\Phi)$. Expanding $V$ to first order over $W$, the part of the full solution linear in $E_{h}$, gives
\begin{equation}
 \delta V[W]=UW+WU+\tau UWU.
 \label{keom:variation}
\end{equation}
Indeed, with $R=(1-\tau\Phi)^{-1}=1+\tau U$, this variation
is $(RWR-W)/\tau$. The first two terms couple additional tree currents to either side of $W$, while the last term couples them to both sides.

Using the Berends–Giele recursions for the two factors in the trial product, the above equation gives the source
\begin{equation}
 \sigma^{(\tau)}_{ak}={} [D_\tau(X_{aj})+D_\tau(X_{j-1,k}) +\tau D_\tau(X_{aj})D_\tau(X_{j-1,k})-D_\tau(X_{ak})] \times J^{(\tau)}_{aj}J^{(\tau)}_{j-1,k}.
 \label{keom:source}
\end{equation}
The extra product of inverse propagators is required by the
higher contacts. The three interaction terms combine through the identity $D_\tau(x)+D_\tau(y)+\tau D_\tau(x)D_\tau(y)=D_\tau(x+y)$. The source therefore vanishes when
$X_{aj}+X_{j-1,k}=X_{ak}$, namely a rectangular zero.

The equation for $W$ is again linear. Iterating the linear equation for the correction sums the
remaining tree diagrams into the outside current, as in
Eq.~\eqref{ceom:outside}.
\begin{equation}
 \A_{\tau,n}=\sum_{a,k}
       \sigma^{(\tau)}_{ak}\J_\tau[P_O],\qquad
 \J_\tau[P]=\frac{\A_\tau[P]}{D_\tau(X_{p_1p_m})},
 \label{keom:splitting}
\end{equation}
with the same ranges and unit convention as in
Eq.~\eqref{ceom:splitting}.

To identify these amplitudes, put $\tau=2i\lambda$. The
propagator obeys
\begin{equation}
 \frac1{D_\tau(X)}=\lambda\cot(\lambda X)-\frac\tau2=e^{-i\lambda X}\frac{\lambda}{\sin(\lambda X)}.
 \label{keom:propagator}
\end{equation}
For the complete classical solution, define
$\Psi=\Phi+\tau V/2$. Since
$\Phi=\Phi_0+D_\tau^{-1}V$ and $V=\Phi^2+\tau\Phi V$,
\begin{equation}
 \Psi=\Phi_0+\lambda\cot\!\bigl(\lambda(-\Box)\bigr)V,\qquad \Psi^2=V-\lambda^2V^2.
 \label{keom:field-change}
\end{equation}
On the branch analytic at the vacuum,
\begin{equation}
 V=\Psi^2+\lambda^2\Psi^4+2\lambda^4\Psi^6+\cdots
   =\sum_{h\geq0}C_h\lambda^{2h}\Psi^{2h+2}.
 \label{keom:catalan}
\end{equation}
These are exactly the propagator and vertices of
Eq.~\eqref{klt-lagrangian}. Both formulations have the same nonlinear source $V$, so their amputated multi-particle currents agree. The resulting lower functions are therefore $A_{2i\lambda}[P]=\F_\lambda[P]$, including at inherited channel values.

Finally, the source coefficient is
\begin{equation}
 D_{2i\lambda}(X_{aj}+X_{j-1,k})-D_{2i\lambda}(X_{ak})
 =e^{i\lambda(X_{aj}+X_{j-1,k}+X_{ak})}
   \frac{\sin[\lambda(X_{aj}+X_{j-1,k}-X_{ak})]}{\lambda}.
 \label{keom:sine}
\end{equation}
The cancellation of three phases follow from the second equality of Eq. \eqref{keom:propagator}. This gives Eq.~\eqref{finite} with
the complete lower functions of Eq.~\eqref{klt-current}.
For the source \(\sigma_{1n}^{(\tau)}\), it contains all external legs except the root $n$. After the root propagator is amputated, there is no outside tree remains so that an identity outside factor is given. The complete coupling tower and
its uniqueness are derived in Appendix~\ref{appendix:interaction-tower}.

\subsection{Pions}
\label{appendix:pion-eom}

The pion derivation uses the same product and source as the cubic
theory, but the equation of motion must be accompanied by a nonlinear
constraint. We use $D=-\Box$ and the normalization
$A_4^\pi=-(X_{13}+X_{24})$. In the minimal parametrization~\cite{KampfNLSM},
\begin{equation}
 U=1-2C+2i\pi,\qquad C=\frac{1-\sqrt{1-4\pi^2}}2,\qquad \mathcal L_\pi=\frac12\Tr\bigl[(\partial\pi)^2+(\partial C)^2\bigr].
 \label{pieom:action}
\end{equation}
The square root denotes the branch analytic at $\pi=0$.
Introduce the auxiliary matrices
\begin{equation}
 \Gamma=\begin{pmatrix}1&0\\0&-1\end{pmatrix},\quad
 F=\begin{pmatrix}C&\pi\\\pi&-C\end{pmatrix},\quad
 Q=\Gamma-2F.
 \label{pieom:matrices}
\end{equation}
The constraint $C-C^2=\pi^2$ and the pion field equation become $Q^2=1,\qquad [Q,DF]=0$.
For example, varying $\mathcal L_\pi-\Tr\Lambda(C-C^2-\pi^2)$
gives $DC=\Lambda-\{C,\Lambda\}$ and
$D\pi=-\{\pi,\Lambda\}$, which imply the equation above.
All matrices retain the block form in Eq.~\eqref{pieom:matrices}.

Let $\pi_{ab}$ and $C_{ab}$ be the ordered coefficients on the
external legs labeling $(a,\ldots,b-1)$, with $\pi_{a,a+1}=1$.
Writing $s_a=(-1)^a$, select the $(s_a,s_b)$ component of $F$:
\begin{equation}
 J_{ab}=\begin{cases}
 \pi_{ab},&b-a\text{ odd},\\
 s_aC_{ab},&b-a\text{ even}.
 \end{cases}
 \label{pieom:currents}
\end{equation}
Expansion of the two pion field equations gives
\begin{align}
 (s_a+s_b)J_{ab}
 &=2\sum_{c=a+1}^{b-1}J_{ac}J_{cb},\label{pieom:algebraic}\\
 (s_a-s_b)X_{ab}J_{ab}
 &=2\sum_{c=a+1}^{b-1}(X_{cb}-X_{ac})J_{ac}J_{cb}.
 \label{pieom:dynamical}
\end{align}
Thus an even number of pion legs is determined algebraically through the above equations, while those with an odd number contain a pion propagator. In particular, the coefficient of the composite field
$C_{ab}$ and the amputated pion current $X_{ab}\pi_{ab}$ depend only on the lower-point currents.
For odd $b-a\geq3$, putting the root on shell gives the complete pion amplitude. The coefficient $C_{ab}$, non-zero for even $b-a$, will be identified below with the
mixed amplitude.

For even $n$, select the external pion $j-1$ and denote its linear
coefficient by $W$. In the following equations $F$ is the
classical solution generated by the remaining external legs. Linearization gives
\begin{equation}
 \{Q,W\}=0,\qquad HW=0,\qquad HW\equiv[Q,DW]-2[W,DF].
 \label{pieom:linearized}
\end{equation}
We use the product proposed in the cubic derivation and define
the source matrix $\Sigma$ through its selected coefficients:
\begin{equation}
 W^{(0)}_{ak}=J_{aj}J_{j-1,k},\qquad \sigma_{ak}=\bigl(X_{aj}+X_{j-1,k}-X_{ak}\bigr) J_{aj}J_{j-1,k}.
 \label{pieom:source}
\end{equation}
The common marked external leg is counted once. Applying
Eqs.~\eqref{pieom:algebraic} and \eqref{pieom:dynamical} to the
two factors verifies both the constraint and the mismatch:
\begin{equation}
 \{Q,W^{(0)}\}_{ak}=0,\qquad \bigl(HW^{(0)}+[Q,\Sigma]\bigr)_{ak} =(s_{j-1}+s_j)(X_{aj}-X_{j-1,k})W^{(0)}_{ak}=0.
 \label{pieom:mismatch}
\end{equation}
The derivation of the second equality is similar as for cubic theory; their
remaining sums are Eq. \eqref{pieom:algebraic}. The cancellation
uses $s_{j-1}=-s_j$. Consequently the correction
$\Delta=W-W^{(0)}$ obeys
\begin{equation}
 \{Q,\Delta\}=0,\qquad H\Delta=[Q,\Sigma].
 \label{pieom:response}
\end{equation}
The correction vanishes when $h$ is the first or last external leg. With these boundary conditions, the linearized recursions Eqs. \eqref{pieom:algebraic}-\eqref{pieom:dynamical} determine each correction from currents with fewer external legs.

To identify the outside factor to each source, introduce a fixed external source coupled to the pion and composite fields,
\begin{equation}
 \mathcal L_\pi[\Sigma] =\frac14\Tr_2(\partial F)^2-\frac12\Tr_2(\Sigma F),\qquad [Q,DF-\Sigma]=0.
 \label{pieom:sourced-action}
\end{equation}
Here $\Tr_2$ includes both matrix traces, and
$\Sigma$ has the block form of $F$, so the source coupling is
$-\Tr(\sigma_\pi\pi+\sigma_C C)$. Diagrammatically, it sums all trees with one source insertion. A pion insertion generates
the complete pion current, including its root propagator. While the insertion of $C$ generates a composite operator form factor without that propagator. After amputating leg \(n\), the remaining diagrams give the outside current on $J[P_O]$ on $P_O=(k,\ldots,n,1,\ldots,a)$. $s_as_k=1$ introduces the phase cancellation from the source and the outside currents. Since the product has no pole in $X_{1n}$,
\begin{equation}
 A_n^\pi =\lim_{X_{1n}\to0}X_{1n}\Delta_{1n} =\sum_{a=1}^{j-2}\sum_{k=j+1}^{n} \sigma_{ak}\,J[P_O],\qquad J[n,1]=1.
 \label{pieom:splitting}
\end{equation}
Every source vanishes on the corresponding rectangular zero, away from poles.

The composite coefficient is the mixed amplitude of
Eq.~\eqref{pionJ}. To see this, use Cayley coordinates, where $C=\varphi^2/(1+\varphi^2)$. In the Cayley-current representation of the single soft theorem, the contribution associated with $k_{m-1}$ requires the last coupled current to be the single external leg $m$. Removing that unit current leaves precisely the tree expansion of $C_{1m}$. Thus, in our normalization and second-trace ordering,
\begin{equation}
 C_{1m}=A_m^{\rm ext}(1,\ldots,m\mid1,m-1,m).
 \label{pieom:mixed}
\end{equation}
This identifies
$J_{ab}=\J^\pi[a,\ldots,b]$ and reproduce Eq.~\eqref{pion} from Eq. \eqref{pieom:splitting}.

\section{Fixing the complete interaction tower}
\label{appendix:interaction-tower}

Within the class of theories in Eq.~\eqref{selection-lagrangian},
the four- and five-point zeros fix the propagator, and one rectangular
zero at each higher multiplicity fixes the remaining contact vertex.

Write $q=1/D$ and $c=g_4/2$, with $g_3=1$. The four-point zero gives
$q(x)+q(-x)=-2c$, so $P(x)=q(x)+c$ is odd. At five points,
substituting $q=P-c$ cancels all terms linear in $P$ and leaves the
five cubic diagrams built from $P$, together with the constant
$\kappa=g_5-5c^2$. On the zero $c_{13}=c_{14}=0$, take
$x=X_{13}$ and $y=X_{24}$. Then $X_{14}=x+y$ and $X_{25}=-x$, and
oddness reduces the zero condition to
\begin{equation}
P(x+y)\bigl[P(x)+P(y)\bigr]=P(x)P(y)-\kappa.
\label{tower:addition}
\end{equation}
Expanding at $y=0$ gives $P'=-P^2-\kappa$. The unit-residue pole
fixes the solution,
\begin{equation}
P(x)=\sqrt{\kappa}\cot(\sqrt{\kappa}\,x),
\qquad q(x)=P(x)-c,
\label{tower:propagator}
\end{equation}
with $P(x)=1/x$ at $\kappa=0$.

For fixed $q$, uniqueness at higher points follows immediately.
At the first valence $m$ where two candidate theories differ, all
exchange diagrams agree. Their amplitudes therefore differ only by
the constant $g_m-g'_m$. Evaluating at any regular point of a common
rectangular zero gives $g_m=g'_m$.

To construct the tower, write $V(\Phi)=\sum_{m\geq3}g_m\Phi^{m-1}$.
For $c=0$, Appendix~\ref{appendix:klt-eom} gives a solution with all
rectangular zeros whose interaction satisfies
$\widehat V=\Psi^2+\kappa\widehat V^2$. For general $c$, set
$\Psi=\Phi+cV(\Phi)$. This turns the field equation
$\Phi=\Phi_0+(P-c)V$ into $\Psi=\Phi_0+PV$. The shift $cV$ has no
root propagator, so it drops out upon amputation. The same tree
amplitudes are therefore generated by
\begin{equation}
\boxed{V=(\Phi+cV)^2+\kappa V^2.}
\label{tower:general-generating}
\end{equation}
Equating powers of $\Phi$ fixes every coupling from $g_3=1$:
\begin{equation}
g_m=2c\,g_{m-1}
 +(c^2+\kappa)\sum_{r=3}^{m-2}g_r g_{m+1-r},
\qquad m\geq4.
\label{tower:general-recursion}
\end{equation}

Only $\kappa$ affects the tree amplitudes. In the convention $c=0$,
Eq.~\eqref{tower:general-recursion} gives the Catalan couplings
$g_{2h+3}=C_h\kappa^h$ and $g_{2h+4}=0$, recovering
Eq.~\eqref{klt-lagrangian} for $\kappa=\lambda^2$. The choice
$c=i\lambda$ instead gives $V=\Phi^2/(1-2i\lambda\Phi)$, the form
used in Appendix~\ref{appendix:klt-eom}.

\section{Proof of one-loop exact splitting}
\label{app:loop-cut-proof}

\paragraph{Cubic scalars.}
We prove Eq.~\eqref{loop} by matching its radial residues at
independent surface kinematics. Cutting $Y_r^+$ opens the punctured
disk into the polygon
\[
Q_r=(1,\ldots,r_-,p,r_+,r+1,\ldots,n),
\]
where $r_\pm$ are the two copies of $r$ and $p$ is the opened
puncture. The residue of the integrand is therefore $A[Q_r]$.

For $r<n$, the cut appears in two places on the right of
Eq.~\eqref{loop}. If $j\leq r<i$, it splits the inner tree along
the diagonal $(r,R)$. The two resulting trees carry the attachments
$Y_j^-$ and $Y_i^+$. Together with the outside current, they give
precisely the tree splitting formula Eq.~\eqref{tree} on a polygon
$\widetilde Q_r$. This is $Q_r$ with the channels $(r_+,h)$,
$1\leq h\leq r$, assigned $Y_h^-$ in place of $X_{rh}$.
Here $h=r$ denotes the copy $r_-$.

If $i=r$, the pole lies in the explicit attachment $1/Y_i^+$.
These terms restore the original channels. Writing $P_T(j,i)=(L,j,\ldots,i,R)$ and
$P_O(i,j)=(i,\ldots,n,1,\ldots,j)$, their sum is
\begin{equation}
\begin{aligned}
A[Q_r]-A[\widetilde Q_r]
={}&\sum_{h=1}^{r}
\left(\frac{1}{X_{rh}}-\frac{1}{Y_h^-}\right)\times A[P_T(h,r)]\,A[P_O(r,h)] .
\end{aligned}
\label{eq:loop-cut-restore}
\end{equation}
To see this, restore the chords $(r_+,h)$ in order of increasing
$h$. Only trees containing the changed chord contribute to each
difference. Cutting that chord gives the two trees displayed above.
The full residue is thus $A[Q_r]$.

For $r=n$, the peak supplies $A[\widetilde Q_n]$ and restoration
starts at $h=2$, since $(n_+,1)$ is a boundary edge. Reflection
gives the $Y_r^-$ cuts. Both sides of Eq.~\eqref{loop} are linear
in each radial propagator separately and vanish when all radial
propagators are set to zero. Hence their residues fix the full
cubic integrand.

\paragraph{Inverse KLT.}
The radial cuts follow by the same argument, using inverse-KLT tree
splitting for the cut inside $P_T$. To see that the endpoint terms
again restore the channels, write $P_\lambda(X)=\lambda\cot(\lambda X)$.
Away from the peak, the sine addition formula gives the coefficient
of the two amputated trees as
\begin{equation}
\begin{aligned}
K_{ij}=
\bigl[P_\lambda(Y_i^+)+P_\lambda(Y_j^-)\bigr]P_\lambda(X_{ij})
-P_\lambda(Y_i^+)P_\lambda(Y_j^-)+\lambda^2 .
\end{aligned}
\label{eq:loop-cut-kernel}
\end{equation}
Thus Eq.~\eqref{eq:loop-cut-restore} carries over with $A$ replaced
by $\F_\lambda$ and $1/X$ by $P_\lambda(X)$. Only the terms
independent of all radial propagators remain to be matched.

Set $\lambda Y_i^\pm=\pi/2$ for every $i$, keeping the boundary
channels arbitrary. The radial propagators and the peak term vanish,
and $K_{ij}=\lambda^2$. Fix a loop contact with $2h$ ordered tree
branches. The inner vertex takes $2a$ consecutive branches and has
weight $C_a\lambda^{2a}$. The outside vertex takes the remaining
$2h-2a$ branches and has weight $C_{h-1-a}\lambda^{2h-2a-2}$.
The branch containing the boundary edge $(n,1)$ must lie outside,
leaving $2h-2a$ placements for the inner group. Including $K_{ij}$,
the total weight is
\begin{equation}
\begin{aligned}
\lambda^{2h}\sum_{a=0}^{h-1}(2h-2a)C_aC_{h-1-a}
=(h+1)C_h\lambda^{2h}
=2(2h-1)C_{h-1}\lambda^{2h}.
\end{aligned}
\label{eq:loop-cut-contact}
\end{equation}
The first equality follows by pairing $a$ with $h-1-a$ and using
the Catalan recursion. This is twice the contact weight of
Ref.~\cite{surfaceKLT}, as required by the two parity sectors.
Odd valence gives no contribution. Since the attached trees were
arbitrary, this proves Eq.~\eqref{loop-sine} including all contacts.

\paragraph{Pions.}
The parity shift preserves $R^{(1)}_{ij}$. The attached tree factors
scale as $O(\delta^{-(|P_T|-2)})$ and $O(\delta^{-(|P_O|-2)})$,
respectively, by Appendix~\ref{appendix:pion-loop}. Since
$(|P_T|-2)+(|P_O|-2)=n$, multiplying the cubic identity by $\delta^n$
and taking the limit gives Eq.~\eqref{pion-loop} term by term.

\section{The lower trees in pion loop splitting}
\label{appendix:pion-loop}

The lower factors in Eq.~\eqref{pion-loop} are complete trees.
Their inherited parity assignments need not alternate, so
ordinary pion amplitudes alone do not specify them. We first
define these trees and then show that only their leading
large-$\delta$ coefficients enter the loop identity.

\subsection{Shifted trees and their insertions}

For $P=(p_1,\ldots,p_m)$, retain the parent signs
$s_{p_a}=\pm1$ and shift each internal channel by
\begin{equation}
 X_{p_ap_b}^{(\delta)}
 =X_{p_ap_b}+\frac\delta2(s_{p_a}+s_{p_b}).
 \label{proj:shift}
\end{equation}
Boundary edges of the lower polygon are not propagators and
are excluded from this assignment. Let $d(P)$ denotes the degree that count its
neighboring external legs and the propagators with equal-sign vertices.
The complete cubic tree obeys
\begin{equation}
A^{(\delta)}[P]
=O\!\left(\delta^{-(m-2-d(P))}\right).
\label{proj:bound}
\end{equation}
For a polygon containing both signs, define
\begin{equation}
\mathscr A[P]
=\lim_{\delta\to\infty}
\delta^{m-2-d(P)}A^{(\delta)}[P].
\label{proj:tree}
\end{equation}
The similarly scaled coefficient of an all-equal-sign polygon
vanishes. Two-vertex units are defined separately and are not
tree amplitudes.

To prove Eq. \eqref{proj:bound}, consider the coefficient of $\delta^{-k}$. Each coefficient has poles only for unshifted propagators joining opposite sign vertices and has degree $k+3-m$ in $X$. On each such pole, tree factorization gives $m_L+m_R=m+2$ and $d_L+d_R=d(P)$. Induction in multiplicity therefore shows that every residue vanishes for $k<m-2-d(P)$. Any remaining contribution must be a local contact term. For $d(P)\geq 1$, its negative momentum degree excludes such a term. For $d(P)=0$, only a possible constant at $k=m-3$ remains. Cyclic symmetry reverses the alternating signs and sends $\delta\rightarrow-\delta$, while this is excluded since $m$ is even. Together with $A_{3}=1$, this proves Eq. \eqref{proj:bound}.

For $d=0$, $\mathscr A$ is the ordinary pion amplitude. An external leg between equal-sign vertices instead carries an insertion of the composite field defined in Appendix \ref{appendix:pion-eom}. Use the same conventions of that appendix, the shifted cubic equation gives
\begin{equation}
 D\Phi+\frac\delta2\{\Gamma,\Phi\}=\Phi^2, \qquad \Phi=\delta F,\qquad Q^2=1+\frac4\delta DF,\qquad [Q,DF]=0.
 \label{proj:two-color}
\end{equation}
At leading order these reduce to the pion constraint and field equation already derived in Appendix \ref{appendix:pion-eom}. For a single composite insertion, choose its external leg as the root. With $p=m-1$ external pions, the corresponding current is $s\delta^{1-p}$ times the coefficient of $C$. Amputation multiplies this by
$X+s\delta$, canceling the sign $s$ and give the scaling $\delta^{2-p}=\delta^{3-m}$ required by Eq. \eqref{proj:tree}. Thus the projected tree is the complete $C$ insertion, whose identification with the mixed amplitude was established in Appendix \ref{appendix:pion-eom}.

The trees with two composite insertions can be generated by adding a scalar probe $S$ to the pion theory,
\begin{equation}
 \Delta\mathcal{L}=\frac12\Tr(\partial S)^2+g\Tr[S C(\pi)].
 \label{proj:probe}
\end{equation}
Take the coefficient of $g^{2}$ in the connected tree amplitude with two external $S$ legs. No internal $S$ propagator occurs at this order. The result includes all pion-exchange diagrams and the contact vertices generated by $C$. By mathematical induction, its factorization residues agree with those of Eq. \eqref{proj:tree}. Their difference therefore has no poles. Since both functions scale as inverse momentum squared for $d=2$, this difference cannot be a local contact term and must vanish. For example,
\begin{equation}
 \mathscr A(S,S,\pi,\pi)=\frac1{X_{24}},\qquad \mathscr A(S,\pi,S,\pi)=\frac1{X_{13}}+\frac1{X_{24}}.
 \label{proj:two-insertions}
\end{equation}
The insertion positions determine the allowed poles. These
functions are not arbitrary amplitudes with four biadjoint
scalars. Alternatively, Eq.~\eqref{proj:tree}, evaluated after
summing the complete cubic tree, is a terminating definition
that requires no auxiliary fields.

\subsection{Projection of each mountain term}

Set $s_i=(-1)^i$, $s_L=+1$, and $s_R=-1$.
The inner and outside polygons are
$P_T=(L,j,\ldots,i,R)$ and
$P_O=(i,\ldots,n,1,\ldots,j)$.
Their equal-sign edge counts are
\begin{equation}
 d_T=\mathbf1_{j\ {\rm even}}+\mathbf1_{i\ {\rm odd}},
 \qquad d_O=\mathbf1_{i\equiv j\ ({\rm mod}\ 2)}.
 \label{proj:counts}
\end{equation}
The mountain numerator is unchanged:
\begin{equation}
 \Delta_\delta(Y_i^++Y_j^--X_{ij}) =\frac\delta2\bigl[(s_i-1)+(s_j+1) -(s_i+s_j)\bigr]=0.
 \label{proj:numerator}
\end{equation}

A propagator with equal-sign endpoints contributes one power
of $\delta^{-1}$; its scaled limit is the common endpoint
sign. An opposite-sign propagator is unchanged. The two
radial attachments therefore supply $d_T$ powers and the
outside attachment supplies $d_O$. Since
$|P_T|+|P_O|=n+4$, Eq.~\eqref{proj:bound} gives
\begin{equation}
\frac{A^{(\delta)}[P_T]}
     {(Y_j^-)^{(\delta)}(Y_i^+)^{(\delta)}}
=O\!\left(\delta^{-(i-j+1)}\right),
\qquad
\J^{(\delta)}[P_O]
=O\!\left(\delta^{-(n-i+j-1)}\right).
\label{proj:attached-bounds}
\end{equation}
Thus each summand starts at order $\delta^{-n}$ or later.
Multiplying Eq.~\eqref{loop} by $\delta^n$ and taking the
limit produces one product of leading coefficients per
summand, with no contributions from subleading coefficients.

Away from the peak $(i,j)=(n,1)$, the projected factors
in Eq.~\eqref{pion-loop} are
\begin{equation}
\begin{aligned}
\left[\frac{A[P_T]}{Y_j^-Y_i^+}\right]_\pi
&=\mathscr A[P_T]
\begin{cases}
(Y_j^-Y_i^+)^{-1},&i\text{ even},\ j\text{ odd},\\
(Y_i^+)^{-1},&i,j\text{ even},\\
-(Y_j^-)^{-1},&i,j\text{ odd},\\
-1,&i\text{ odd},\ j\text{ even},
\end{cases}\\
\left[\J[P_O]\right]_\pi
&=\mathscr A[P_O]
\begin{cases}
X_{ij}^{-1},&i\not\equiv j\pmod2,\\
(-1)^i,&i\equiv j\pmod2.
\end{cases}
\end{aligned}
\label{proj:factors}
\end{equation}
The peak $(i,j)=(n,1)$ is treated separately:
$[\J[n,1]]_\pi=1$, and its inner polygon is alternating.
Only the inner tree can have two insertions; the outside
tree has at most one.

This proves Eq.~\eqref{pion-loop} for the shifted surface
representative, retaining its tadpole and bubble terms.
The result concerns the integrand before integration. The
auxiliary tree prescription does not impose a choice of
subtractions for a conventional loop amplitude.

\section{Pion soft limits from exact splitting}
\label{appendix:soft}

We derive the Adler zero and its first nonzero coefficient. We use the normalization
$A_4^\pi=-(X_{13}+X_{24})$ and take smooth on-shell,
momentum-conserving soft limits at generic hard kinematics.
Hard momenta in the leading coefficients below are evaluated at
zero soft momentum.

The limit follows from $j=3$ splitting. Since
$\J^\pi[1,2,3]=-1$, Eq.~\eqref{pion} becomes
\begin{equation}
 A_n^\pi=-\sum_{k=4}^{n}\left(\sum_{i=3}^{k-1}c_{1i}\right)
 \J^\pi[2,\ldots,k]\J^\pi[k,\ldots,n,1],\qquad \J^\pi[n,1]=1.
 \label{soft:split}
\end{equation}

Set $k_1=\varepsilon q$. Every numerator in
Eq.~\eqref{soft:split} is $-2\varepsilon q\cdot\sum_{i=3}^{k-1}k_i$.
The lower factors are finite: every pion propagator carries a
proper odd interval of hard waves with generically nonzero
invariant. The closing channel $X_{2n}$ tends to zero, but
$\J^\pi[2,\ldots,n]$ is a composite current with no root propagator.
Thus the Adler zero holds term by term. Interchanging the two
finite sums also determines the first nonzero coefficient:
\begin{equation}
 A_n^\pi(\varepsilon q,k_2,\ldots,k_n)
 =2\varepsilon\sum_{i=3}^{n-1}(q\cdot k_i)
 \sum_{k=i+1}^{n}\left[\J^\pi[2,\ldots,k]\J^\pi[k,\ldots,n,1]\right]_{k_1=0}
 +O(\varepsilon^2).
 \label{soft:single-currents}
\end{equation}

The inner sum is a complete mixed amplitude, with the second
trace ordered as $(2,i,n)$:
\begin{equation}
 \sum_{k=i+1}^{n}\left[\J^\pi[2,\ldots,k]\J^\pi[k,\ldots,n,1]\right]_{k_1=0}
 =A_{n-1}^{\rm ext}(2,\ldots,n\mid2,i,n).
 \label{soft:mixed-current}
\end{equation}
To verify this identity, use the minimal fields and Cayley
currents mentioned in Appendix~\ref{appendix:pion-eom}.
At $k_1=0$, the product on the left is $C_{2k}\pi_{k1}$ for
even $k$ and $-\pi_{2k}C_{k1}$ for odd $k$. The fields obey
\begin{equation}
 \pi=\frac{\varphi}{1+\varphi^2}
 =\sum_{r\geq0}(-1)^r\varphi^{2r+1},\qquad
 C=\frac{\varphi^2}{1+\varphi^2}
 =\sum_{r\geq1}(-1)^{r-1}\varphi^{2r}.
 \label{soft:cayley-fields}
\end{equation}
Fix a partition $(2,\ldots,n)=B_1\cdots B_{2r+1}$ into
nonempty odd-length blocks, and let $i\in B_t$. A cut after
block $\ell$ contributes $(-1)^{r-1}(-1)^\ell\prod_a K(B_a)$.
The condition $k>i$ permits precisely $\ell=t,\ldots,2r$, and
\begin{equation}
 \sum_{\ell=t}^{2r}(-1)^\ell=
 \begin{cases}1,&t\text{ even},\\0,&t\text{ odd}.
 \end{cases}
 \label{soft:alternating-cuts}
\end{equation}
Consequently the sum retains exactly the partitions in which
$i$ belongs to an even block, with weight $(-1)^{r-1}$.
These are the Cayley vertex weights for a three-scalar junction,
with scalar branches containing $2,i,n$ and cubic seed
$A_3^{\rm ext}=1$~\cite{LowYin2018}. Each branch equals $K(B)$:
every vertex with two scalars has the corresponding pion vertex
rule, so replacing its scalar path by pions preserves the current.
A tree with three external scalars has exactly one such junction.
This proves Eq.~\eqref{soft:mixed-current}, including contact
terms. Substitution gives the subleading single-soft theorem
\cite{extensions}:
\begin{equation}
 A_n^\pi(\varepsilon q,k_2,\ldots,k_n)
 =2\varepsilon\sum_{i=3}^{n-1}(q\cdot k_i)
 A_{n-1}^{\rm ext}(2,\ldots,n\mid2,i,n)+O(\varepsilon^2).
 \label{soft:single-theorem}
\end{equation}

\end{document}